\documentclass[twocolumn,times]{aastex63}

\usepackage{enumitem}

\submitjournal{ApJ}

\shorttitle{Search for High-energy Neutrinos from TXS0506+056}
\shortauthors{Aab et al., The Pierre Auger Collaboration}

\begin{document}


\title{A search for ultra high energy neutrinos from TXS 0506+056 using the Pierre Auger Observatory}


\begin{abstract}
Results of a search for ultra-high-energy neutrinos with the Pierre
Auger Observatory from the direction of the blazar TXS 0506+056 are presented.
They were obtained 
as part of the follow-up that stemmed from the detection of high-energy neutrinos and gamma rays with IceCube, \textit{Fermi}-LAT, MAGIC, and other 
detectors of electromagnetic radiation in several bands. 
The Pierre Auger Observatory is sensitive to neutrinos in the energy range
from 100 PeV to 100 EeV and in the zenith angle range from $\theta=60^\circ$ to
$\theta=95^\circ$, where the zenith angle is measured from the vertical direction. No neutrinos from the direction of TXS 0506+056 have been found. 
The results were analyzed in three periods: One of 6 months around the
detection of IceCube-170922A, coinciding with a flare period
of TXS 0506+056, a second one of 110 days during which the
IceCube collaboration found an excess of 13 neutrinos from a direction
compatible with TXS 0506+056, and a
third one from 2004 January 1 up to 2018 August 31, over which the Pierre Auger Observatory
has been taking data. 
The sensitivity of the Observatory is addressed for different spectral 
indices by considering the fluxes that would induce a single expected event during the 
observation period. For indices compatible with those measured by the IceCube 
collaboration the expected number of neutrinos at the Observatory is well-
below one. 
Spectral indices as hard as 1.5 would have to apply in this energy range to expect 
a single event to have been detected. 
\end{abstract}




\section{Introduction}

On 2017 September 22, a through-going muon that deposited 23.7 TeV 
was detected at the IceCube telescope~\citep{IceCube} in  
Antarctica, likely to be produced by a neutrino. The most probable 
energy for the neutrino is 290 TeV~\citep{Multimessenger} assuming a spectrum compatible 
with the measured diffuse flux~\citep{Aartsen:2015knd,Aartsen:2016xlq}. 
Within a minute, the arrival direction and 
energy estimates were reported through the Gamma-ray
Coordinates Network Circular~\citep{GCNAlert} as part of the 
routine of the multi-messenger that is being established in 
high-energy astrophysics. 
Six days later, flaring activity in the $\gamma$-ray band was observed 
in the Fermi Large Area Telescope (\textit{Fermi}-LAT), 
from TXS 0506+056 (RA= 9h 55.6m, Dec= +5$^\circ$ 41.6'), 
a powerful blazar at relatively high redshift of $0.3365\pm 0.0010$~\citep{zOfTXS} and only $0.1^\circ$ away from the deduced neutrino direction~\citep{FermiLATTXS}. 
The chance possibility of this correlation was estimated to be $\sim 0.3 \%$ (3$\sigma$ level) which motivated further scrutinizing of this object in
practically all bands of the electromagnetic spectrum and in
neutrinos~\citep{Multimessenger,Padovani_TXS}. 

The remarkable multi-messenger effort that followed revealed a complex 
variable activity. Most notably, the search of 
archival IceCube data for signal correlations from the same direction also revealed 
an excess flux of $13\pm5$ through-going muons between December 2014 and February 2105, 
dominating the background neutrino flux from this region, which was 
interpreted as a burst of neutrinos over a time window of about 110 days
from the same object~\citep{archivalIce3Search}. The significance of such 
an excess localized in the reported time window being due to background atmospheric neutrinos  
is estimated to be at the $3.5\sigma$ level. A Very-High-Energy (VHE) signal between 
80 and 400 GeV was detected in the MAGIC telescope when integrating observation between  
2017 September 24 and 2017 October 4, confirming the flaring activity observed by \textit{Fermi}-LAT~\citep{MAGIC_TXS}.  

The detection of neutrinos from the direction of TXS 0506+056 illustrates the potential of multi-messenger observations~\citep{Multimessenger}. 
The follow-up studies have attracted a lot of attention since the detection of neutrinos from blazars would provide the first robust evidence of hadronic acceleration in astrophysical jets, potentially explaining the diffuse high-energy 
neutrino excess detected by IceCube over the atmospheric background,  
and providing much insight into the modeling of these powerful objects. 
While the energy of the gamma rays that can reach us from such an extragalactic source is 
limited by pair production in the background radiation fields, neutrinos are not 
prone to similar interactions and will travel unimpeded up to the highest energies. 
Moreover, TXS 0506+056 is listed among the 50 brightest objects in the third catalog of active galactic nuclei detected by \textit{Fermi}-LAT \citep{3rdFermi-LAT-catalogue}, suggesting that neutrino emission may be highly non-uniform within the blazar population~\citep{KeivaniHalzen19}. 
Naturally, other neutrino searches have followed with other 
facilities spanning different energy bands, such as ANTARES~\citep{AntaresTXS} and 
Kamiokande~\citep{KamiokandeTXS}. These searches have reported no signals. 

Ultra-High-Energy (UHE) neutrinos have been searched for with the Pierre Auger Observatory 
since 2004~\citep{AugerNuTau2008} by looking for inclined showers that
develop deep in the atmosphere. The Observatory has been shown to have a similar 
sensitivity to that of IceCube for UHE neutrinos of energies above 
100 PeV~\citep{Auger_nus_JCAP2019}. 
Moreover, it has been shown to have  a distinctive directional sensitivity which 
can have a unique potential to search for transient events from point sources 
that are at preferred declinations~\citep{Auger_nus_point_JCAP2019}. 
This is partly due to the enhanced capability of the Observatory to trigger on air showers
produced by the decay of tau leptons originating from Earth-skimming
tau neutrino interactions near the surface of the Earth. 
The search for correlated neutrinos
from the direction of TXS 0506+056 with the Pierre Auger Observatory has lead
to a negative result~\citep{AugerNeutrinosICRC2019}. In this article we 
describe the search made and the
implications for the possible neutrino flux that could be emitted from this
object in the UHE band. 

\section{The search of neutrinos with the Pierre Auger Observatory}

Ultra-high-energy neutrinos arriving with high zenith-angles can induce air showers deep in the atmosphere. These can be detected with
arrays of particle detectors, such as those operated within the Pierre Auger Observatory located in the Mendoza province, 
Argentina. This Observatory is 
the largest and highest-precision detector available to measure cosmic rays of EeV energies 
and above~\citep{PierreAugerObs}. It consists of a 3000~km$^2$ array 
of water-Cherenkov detectors, the Surface Detector (SD), at ground level
arranged on a triangular grid with 1600 detectors 1.5~km apart. 
The SD samples the front of the extensive air- 
showers that develop when UHE cosmic rays interact in the upper layers of the atmosphere. 
The Pierre Auger Observatory also includes a Fluorescence Detector (FD) 
comprising 27 telescopes that are used to view the atmosphere over the array and capture
the fluorescence light that is emitted as the shower passes through the atmosphere. 
The Observatory was designed to detect cosmic-ray showers, produced by nuclei
or protons interacting in the upper layers of the atmosphere. 

When regular cosmic-rays (i.e.\ protons, nuclei, photons) arrive with zenith angles exceeding about $60^\circ$, their induced showers are largely absorbed in the atmosphere, well-before reaching ground 
level. As a result, the shower front at observation level is mostly composed of ultra-relativistic muons that 
have small time spreads, giving  
characteristic sharp signals in the particle detectors of the array. 
Neutrinos, on the other 
hand, can interact deeper in the atmosphere so that when the neutrino-induced shower 
front reaches ground level it still has a large fraction of electrons, positrons, and 
photons (the electromagnetic component), which gives a signal in the detectors which 
is typically distributed over a larger time interval~\citep{Capelle_1998}. For instance, while an $80^\circ$ proton shower has a signal spread of $\sim 100$~ns about 1~km from the shower axis, 
a neutrino shower can reach over $1~\mu$s. 

The identification of the electromagnetic signals in the SD stations provides the basis 
for the discrimination between neutrino-induced showers and those from the
background of hadronic cosmic-rays. 
This is basically done using variables related to the ratio of the integrated 
signal to its peak value in selected stations which provide a measure of the time-width of 
the signal. For optimization purposes the search is carried out using three distinct groups of
events. Each group is selected from the data with its own set of criteria,
based on variables that relate to the zenith angle and broadly correspond 
to zenith angles between $60^\circ$ and $75^\circ$, ``Downward-Going Low'' (DGL), those 
between $75^\circ$ and $90^\circ$, ``Downward-Going High'' 
(DGH), and those between $90^\circ$ and $95^\circ$, ``Earth-Skimming'' (ES).  
The method devised for each group, in addition to the selection of events, includes the search for
neutrino candidates within, as explained in detail in
\citep{Auger_nus_PRD2015,Auger_nus_JCAP2019}. The ES search provided the most
competitive limit for diffuse flux of UHE neutrinos in the 100~PeV to 10~EeV
range in 2008~\citep{AugerNuTau2008}, before IceCube had been fully 
completed. Later updates including the DGH and DGL searches have set 
limits~\citep{Auger_nus_PRD2015,Auger_nus_JCAP2019} comparable to contemporary 
bounds from IceCube in the same energy range~\citep{IceCube_PRD2018}. 

For the bounds reported here, we make the implicit assumption that the fluxes of 
all neutrino flavors are equal because of flavor oscillations as they travel
to Earth~\citep{Learned_APP1995}, corresponding to a flavor ratio of $\nu_e:\nu_\mu:\nu_\tau$ of 1:2:0 at the source. It is however possible that the actual flavor ratios at the source are significantly different, as has 
been for instance argued in case of acceleration of secondaries in flares~\citep{Klein:2012ug,Winter:2014tta}. 
These could result in modified flavor ratios at Earth that would require reevaluation.
The remarkable sensitivity of the Observatory
to UHE neutrinos is in part due to the Earth-skimming channel~\citep{Bertou_APP2012}, 
in which tau neutrinos traveling through the Earth interact just below the
surface, producing a tau lepton that exits to the atmosphere and induces an
upcoming air shower. Because of this channel, the Pierre Auger 
Observatory provides complementary information to IceCube relative to the tau flavor. 

\begin{figure*}
\begin{center}
\resizebox{0.75\textwidth}{!}{\includegraphics{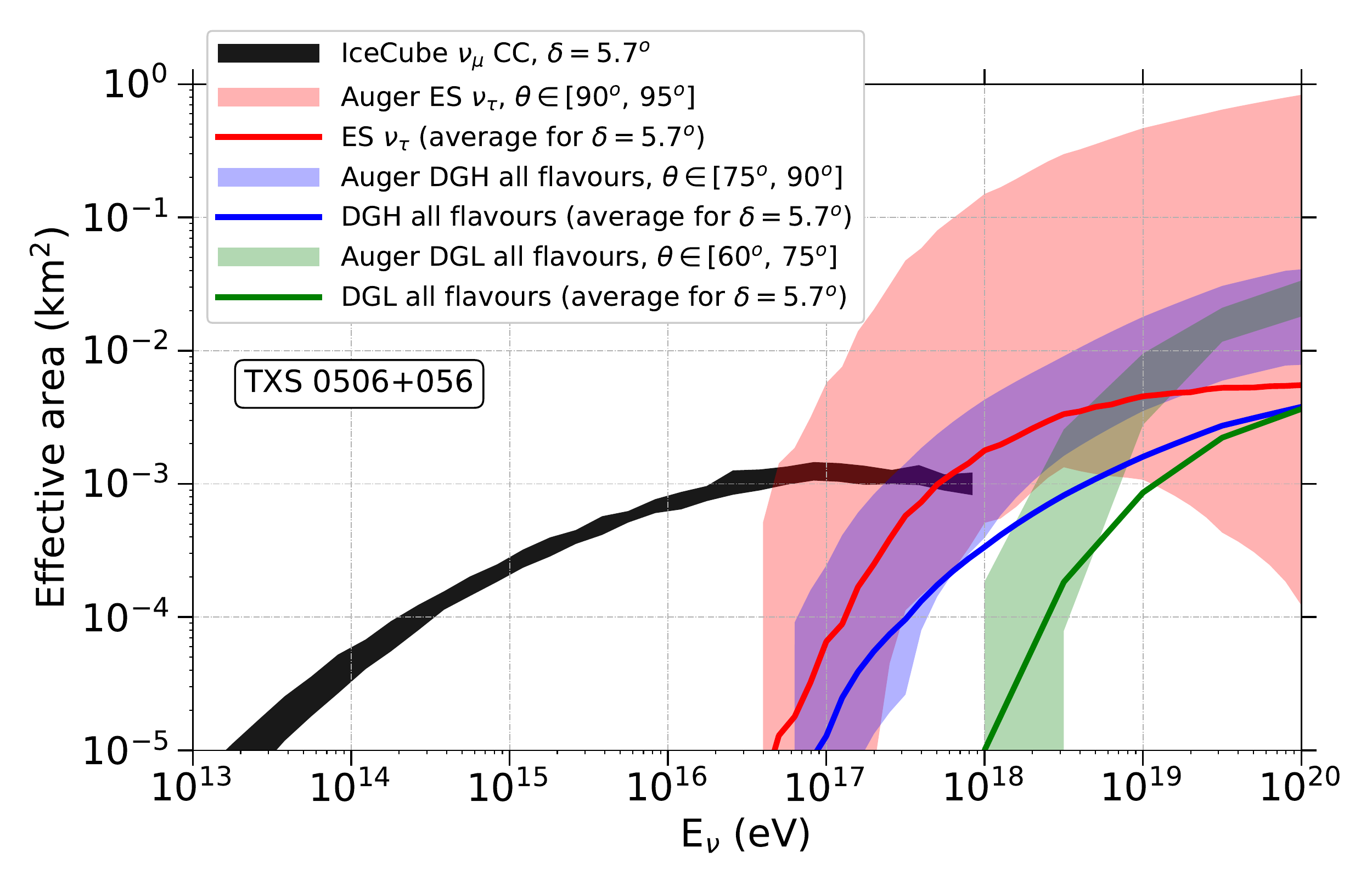}}
\end{center}
\caption{Effective area of the Pierre Auger Observatory as a
  function of neutrino energy for each search channel. The shaded bands bound the  instantaneous effective area for each neutrino detection channel and indicate the 
  variation with zenith angle in the corresponding range. TXS 0506+056 at a
  declination $\delta\simeq5.7^\circ$ is viewed at the SD of Auger for a
  limited amount of time (see Fig.\,\ref{fig:visibility}) and with a range of zenith angles from $\theta=60^\circ$ to $\theta=95^\circ$, the sensitivity being largest below
  the horizon ($\theta>90^\circ$). The full lines represent the effective area
for the different detection channels when averaging over a full day, i.e.\ when including the periods during a day, when the source cannot be seen. The instantaneous effective area of
  IceCube for the declination of TXS 0506+056 is also shown for comparison. 
  For IceCube at the South Pole the zenith angle of TXS 0506+056
  is practically constant over time and given by $\theta = 90^\circ + \delta$. 
  The width of the grey band corresponds in this case to 
  different stages of IceCube construction and configuration which depend on
  the period under consideration.}
\label{fig:EffectiveAreas}
\end{figure*}

There are a number of coincidences 
that make this channel most effective. The matter depth of the Earth's 
chord is a rapidly varying function as the nadir angle of the upcoming tau
neutrino, $180^\circ-\theta$, 
approaches the horizontal ($\theta=90^\circ$). Depending on this 
angle there is a characteristic neutrino energy, $E_{\mathrm{ch}}(\theta)$, at which the 
matter depth matches the neutrino mean free path. This angle roughly optimizes 
the search for neutrinos of energy $\sim E_{\mathrm{ch}}(\theta)$. On the other hand, for 
the SD to detect these showers, they must be nearly horizontal so that the 
shower develops at a very low altitude and the shower front reaches the ground 
as it extends laterally. It turns out that for nadir angles between $85^\circ$
and $90^\circ$, the values of $E_{\mathrm{ch}}(\theta)$ are in the 100 PeV to 10 EeV range, 
large enough to induce showers that can be detected by such a sparse array. 
Moreover, at about 1 EeV, there is a sweet-spot in which the probability for the tau neutrino to convert and for the tau lepton to exit the Earth and to be detected in the SD is  maximal~\citep{NuTauSim_2018}. This is because of a combination of different effects:
Up to about 1 EeV, the matter depth the tau lepton is able to traverse before exiting the Earth is mainly governed by the tau decay length, which increases linearly with energy, enhancing the effective detector volume.
Above about 1 EeV, energy losses in the Earth start to dominate in the exit path of the tau lepton so that its range only rises logarithmically with energy. In addition, because of the increased tau decay length in the atmosphere, showers more often start developing at higher altitudes making the detection of the shower front by the surface detector array less likely~\citep{Zas_NJP2005}.

As a result, for the Earth-skimming detection, the neutrino arrival directions must 
be within a very small angular range of few degrees below the horizon. For 
these directions, the effective area of the Observatory for detecting tau-flavor 
neutrinos is very much enhanced relative to the search method for downward-going 
neutrinos (DGH and DGL). This is the reason why the Pierre 
Auger Collaboration could set the best limit to UHE neutrinos from GW170817~\citep{GW170817_BNS_nus}, the binary neutron star merger event 
detected in gravitational waves and followed up in most bands of the 
electromagnetic spectrum~\citep{GW170817_BNS}. 
The instantaneous effective area is highly 
dependent on the arrival zenith-angle which is a function of the source declination 
and the hour-angle, so that the sensitivity of the Observatory  is highly directional and time-dependent~\citep{Auger_nus_point_JCAP2019}. 
This can be appreciated in Fig.~\ref{fig:EffectiveAreas} where the three wide coloured bands 
span the instantaneous effective area of the Observatory within the zenith angle intervals corresponding to the three search channels. 
For the  Earth-skimming channel the width is largest, reflecting the rapid variation of effective area as the zenith angle changes by only 5 degrees from $90^\circ$ to $95^\circ$ reaching a maximum at $\sim 91^\circ$. 

\begin{figure*}
\begin{center}
\resizebox{0.75\textwidth}{!}{\includegraphics{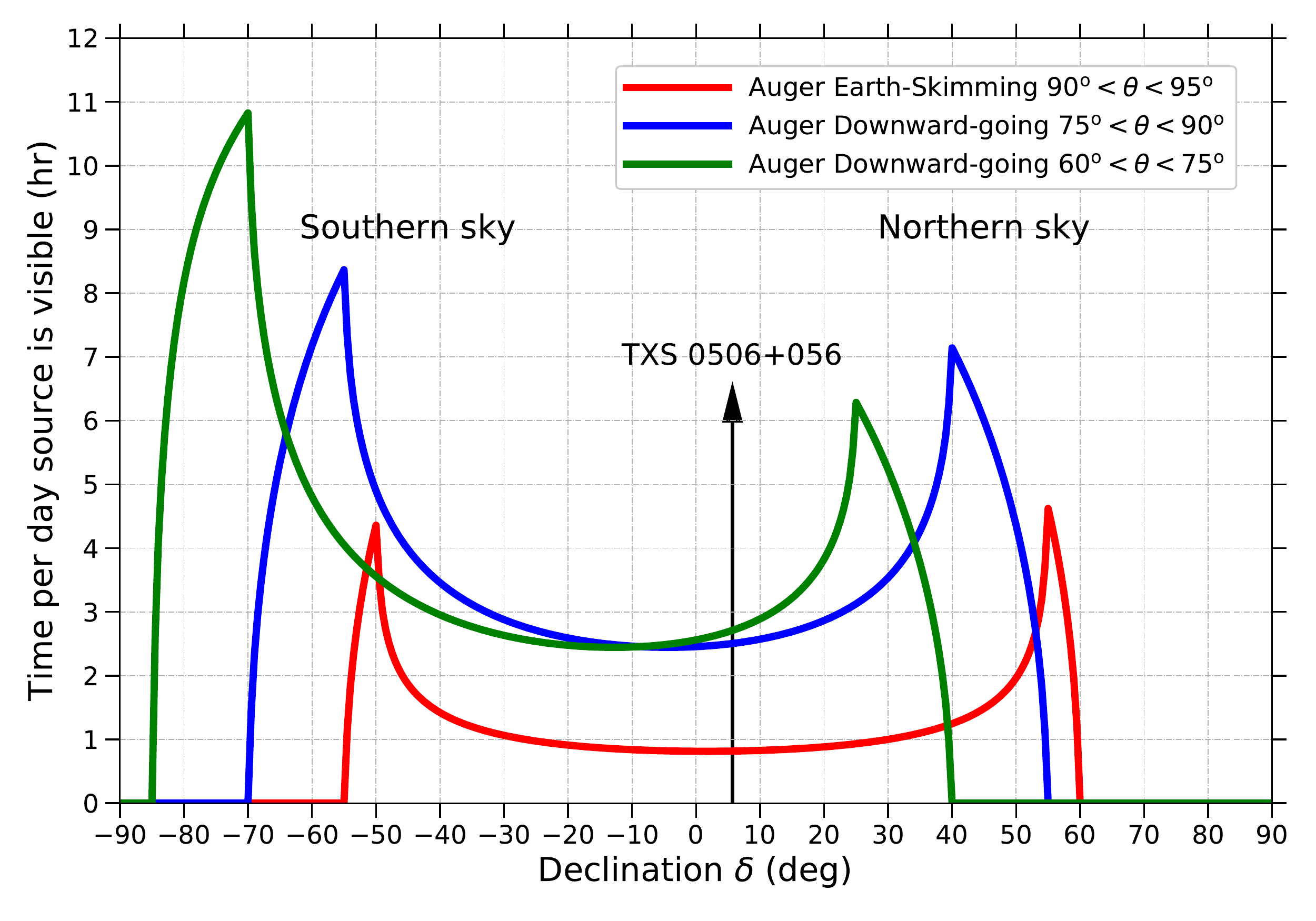}}
\end{center}
\label{fig:visibility}
\caption{Hours per day a source is visible in each of
the search channels as a function of declination. The declination of TXS
0506+056 is marked with an arrow.}
\end{figure*}

The search for neutrinos from the direction of TXS 0506+056 will
be considered for periods much longer than a day. Thus, the effective 
area for neutrino detection must be integrated over time as the source position transits 
over different zenith angles. In Fig.~\ref{fig:EffectiveAreas} we have also shown the daily 
average of the effective area for the Observatory in each of the three search channels for the blazar declination of $5.7^\circ$ (full colored lines), where they are compared to the effective area of IceCube detector for the same source~\citep{archivalIce3Search}. Due to the location of the IceCube detector, the effective area for a fixed position in space depends only
on its declination and is otherwise independent of time for each configuration. The width of the IceCube band here is due to the different configurations
achieved after different construction stages \citep{archivalIce3Search}. 
The effective exposure can be approximately calculated by multiplying the daily average of 
the effective area for the corresponding declination, by the length of the time period under 
consideration~\citep{Auger_nus_point_JCAP2019}. The daily average depends strongly on 
declination and this is partly because the source is only ``visible'' in neutrinos during a varying 
fraction of the day in each zenith angle range. This fraction is displayed in Fig.~\ref{fig:visibility} as 
a function of the declination for each of the three types of searches. The black arrow marks the 
declination of TX0506+056, indicating that the source is not at a declination that maximizes the 
observation time. This effect also contributes to the large variations in effective area as a function of 
the source declination. For periods much larger than a sidereal day the approximation is very
accurate because variations in effective area with time have been relatively small since the 
Observatory was completed in 2008 June. 

\section{Results and discussion}

\begin{figure*}
\begin{center}
\resizebox{0.75\textwidth}{!}{\includegraphics{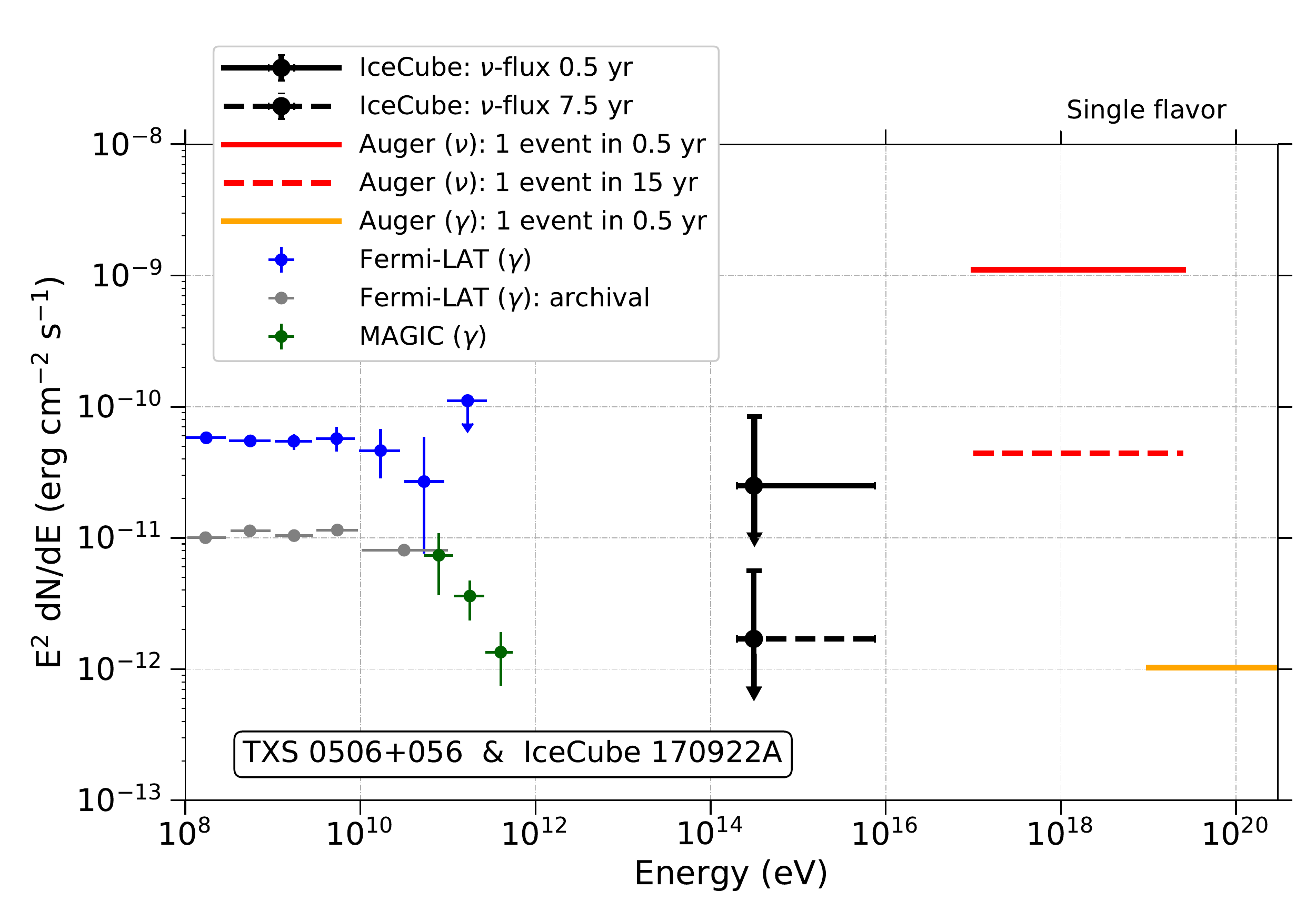}}
\end{center}
\caption{UHE flux reference that would give one
expected neutrino event at the Pierre Auger Observatory over a period of half
a year (2017 March 22 - September 22) for a spectrum $dN/dE \propto E^{-2}$ in comparison to the flux that would 
produce on average one detection like the IceCube-170922A event over the same period (solid red and black lines). 
Flux references are also shown for the Pierre Auger Observatory for a period 
of $\sim$15 years during which it has taken data (2004 January 1 - 2018 August 31) and for a period of 
7.5 years for IceCube~\citep{Multimessenger} (dashed red and black lines). 
The average VHE and UHE photon fluxes measured with \textit{Fermi}-LAT, 
and MAGIC 
around 2017 September 22~\citep{Multimessenger}, and the archival photon measurement from \textit{Fermi}-LAT~\citep{Fermi-Lat-Acero-Archival}, as well as 
the UHE photon flux from this direction that would give one expected photon event in half a year at the Pierre Auger Observatory, are also shown 
for comparison.}
\label{fig:limits}
\end{figure*}
All the data collected with the Pierre Auger Observatory were searched for candidate neutrino events in 
the direction of TXS 0506+056 with 
negative results. Instead of providing a flux limit we calculate the expected flux 
that would have been deduced if a single neutrino had been observed, assuming
a steady flux over a given period of time. This illustrates the expected 
sensitivity to a given flux and can be easily 
converted to a flux limit at $90\%$ confidence multiplying it by a factor 
$2.39$~\citep{Feldman-Cousins}. The results 
naturally depend on the assumptions that are made with respect to the 
time period over which the search is integrated. Two benchmark scenarios have 
been discussed in the original article addressing the correlated detection in 
neutrinos and in the HE and VHE gamma-ray bands~\citep{Multimessenger}.  
The first is of half a year and it is motivated by 
the time window that gave the largest
significance to a search for an excess of neutrino-compatible events in the
archival data of IceCube, interpreted as a neutrino flare~\citep{archivalIce3Search}. 
The second period corresponds to 7.5 years, the whole observation time that 
the IceCube detector had been in operation at the time of detection. 
We here address similar scenarios of half a year and the whole observation period  
of the Pierre Auger Observatory which is 15 years from 2004 January 1 to
2018 August 31. We note that periods over which the SD was unstable have been 
removed from the analysis and that during the first four years of operation the effective area 
was a rapidly growing function of time because the Observatory was under 
construction until 2008 June. 

The average spectral fluxes of UHE neutrinos with a fixed spectral index 
($\sim E^{-\gamma}$) that would produce a 
single event at the Observatory for these two periods 
are displayed in Fig.~\ref{fig:limits} for a spectral index of $\gamma =2.0$, 
assumed to hold in the energy range between 100 PeV and 10 EeV and 
to be constant in time during the corresponding time period. 
In this plot they are compared to the fluxes obtained from the neutrino 
detected in 2017 September 22 and inferred to have energy of order few hundred
TeV, considering a period of half a year and 7.5 years. 
The plot also displays the average VHE gamma-ray flux detected with 
\textit{Fermi}-LAT and MAGIC over periods within a couple of weeks 
around the neutrino detection date of 2017 September 22~\citep{Multimessenger}. 
These gamma-ray fluxes correspond to the reported flaring activity and 
have not been corrected for absorption in the extragalactic background light. They are considerably larger than the average 
gamma-ray fluxes that had been recorded to date from this source, and which 
are also illustrated for comparison~\citep{Fermi-Lat-Acero-Archival}. The sensitivity of the Auger Observatory to 
UHE neutrinos is about an order of magnitude below extrapolations with
$E^{-2}$ spectra, partly due to the non optimal position of the source.

\begin{figure*}
\begin{center}
\resizebox{0.75\textwidth}{!}{\includegraphics{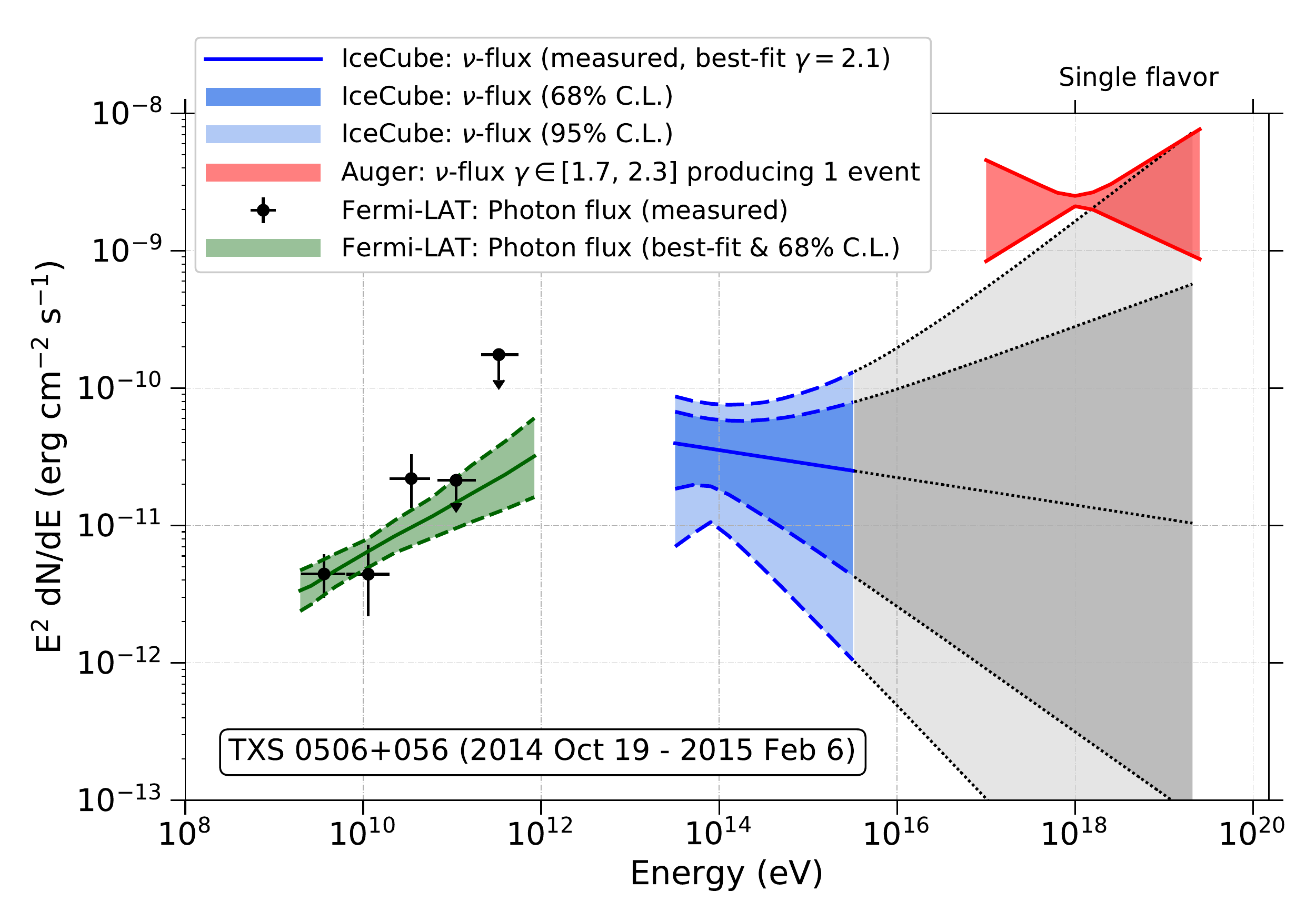}}
\end{center}
\caption{UHE neutrino flux sensitivities for the Pierre Auger
Observatory (one event expected) assuming a constant flux during a period of 
110 days from 19 October 2014 to 6 February 2015 in comparison to the measured 
photon flux~\citep{Padovani_TXS} and to the neutrino flux inferred with IceCube during the same
period with a spectral index of $\gamma=2.1 \pm 0.2$~\citep{archivalIce3Search}. 
The band shown for IceCube is obtained using the extreme values of $\gamma$ 
($\sim$1.75,~$\sim$2.45) from the given one-sigma contour plot and 
($\sim 1.5$ and $\sim 2.7$) from the two-sigma 
contour.} 
\label{fig:limits_flare}
\end{figure*}

We have also compared the sensitivity of the Observatory to the neutrino flux
observed by IceCube between 19 October 2014 and 6 February 
2015. The analysis of this
period resulted in constraints for the normalization and spectral index of the
observed fluence~\citep{archivalIce3Search}. This period of increased neutrino flux in IceCube was not coincident with a
VHE gamma-ray flare from the same source, although a hardening of the spectrum
in the GeV region was reported~\citep{Padovani_TXS}. In
Fig.~\ref{fig:limits_flare} we display the $1\sigma$ and $2\sigma$ bands of
the average flux obtained from the fluence reported assuming an activity
period of 110 days as obtained from the IceCube data analysis using a Gaussian
window. The bands are calculated using the whole parameter space allowed at 
$68\%$ and $95\%$ confidence levels in the IceCube analysis.
The extreme
values of the spectral index are $\gamma \sim 1.75$ and 
$\gamma \sim 2.45$ ($\sim 1.5$ and $\sim 2.7$) for the $68\%$ ($95\%$) CL 
contour plot~\citep{archivalIce3Search}. The figure also displays the average gamma-ray flux obtained for this period illustrating the reported hardening~\citep{Padovani_TXS}. 
The results obtained indicate that the
Pierre Auger Observatory could only be expected to have detected a signal if
the flux extrapolated to the EeV regime with spectral indices 
harder than $\gamma \sim 1.5$.

With the Pierre Auger Observatory it is also possible to search for UHE 
photons~\citep{AugerPhotonsAPJL,AugerPhotonsICRC17,AugerMM19}.
For a source as distant as TXS 0506+056, any UHE photon flux that could have been produced is expected to be strongly attenuated through interactions with the cosmic photon-background fields, unless new physics would occur. 
The data have been searched for UHE photons between 10 and 300 EeV in
coincidence with IceCube-170922A over a period of half a year and also in
coincidence with the 110 day interval interpreted by \citep{archivalIce3Search} as burst of neutrinos.
No event has been found with an angular distance to the source below $2^\circ$. The shower with closest angular distance to the source ($2.1^\circ$) was observed for the latter period and the corresponding value of the Principal Component (PC) for photon discrimination~\citep{MoreAugerPhotons} is very low, so that less than $0.1\%$ of the simulated photons have a smaller PC value. As a result, the probability of this event to be a correlating photon is less than $6 \times 10^{-5}$. Assuming an $E^{-2}$ spectrum, the photon energy flux that would give one expected photon event at the Observatory is 
$1.8 \times 10^{-12}$~erg~cm$^{-2}$~s$^{-1}$.
For the half a year period in 2017 the closest event, at an angular distance of $3.0^\circ$, has an even lower probability to be a
photon, and the reference energy flux for one detected photon becomes $1.0 \times 10^{-12}$~erg~cm$^{-2}$~s$^{-1}$. 

In summary, we have studied the implications of the non-observation of UHE neutrinos with 
the Pierre Auger Observatory. The source is not located at one of the
preferential declinations for observation so the flux constraints that can be 
obtained are rather limited. 
The neutrino flux from TXS 0506+056 at 100s of TeV sampled by IceCube with 
event IceCube-170922 was converted to a flux using a half a year 
period~\citep{Multimessenger}. If the flux from the source had an $E^{-2}$ 
spectrum extending to the EeV and if it had remained constant over the 
lifetime of the Observatory with the same normalization, one neutrino event
could be expected to have been observed at the Pierre Auger Observatory. 
We have also shown that the Observatory could have a chance to detect UHE
neutrinos produced between October 2014 and February
2015 only in case the spectrum extended to the EeV range with a spectral index 
harder than $\gamma \sim 1.5$~\citep{archivalIce3Search}.


\section*{Acknowledgments}

\begin{sloppypar}
The successful installation, commissioning, and operation of the Pierre
Auger Observatory would not have been possible without the strong
commitment and effort from the technical and administrative staff in
Malarg\"ue. We are very grateful to the following agencies and
organizations for financial support:
\end{sloppypar}

\begin{sloppypar}
Argentina -- Comisi\'on Nacional de Energ\'\i{}a At\'omica; Agencia Nacional de
Promoci\'on Cient\'\i{}fica y Tecnol\'ogica (ANPCyT); Consejo Nacional de
Investigaciones Cient\'\i{}ficas y T\'ecnicas (CONICET); Gobierno de la
Provincia de Mendoza; Municipalidad de Malarg\"ue; NDM Holdings and Valle
Las Le\~nas; in gratitude for their continuing cooperation over land
access; Australia -- the Australian Research Council; Brazil -- Conselho
Nacional de Desenvolvimento Cient\'\i{}fico e Tecnol\'ogico (CNPq);
Financiadora de Estudos e Projetos (FINEP); Funda\c{c}\~ao de Amparo \`a
Pesquisa do Estado de Rio de Janeiro (FAPERJ); S\~ao Paulo Research
Foundation (FAPESP) Grants No.~2019/10151-2, No.~2010/07359-6 and
No.~1999/05404-3; Minist\'erio da Ci\^encia, Tecnologia, Inova\c{c}\~oes e
Comunica\c{c}\~oes (MCTIC); Czech Republic -- Grant No.~MSMT CR LTT18004,
LM2015038, LM2018102, CZ.02.1.01/0.0/0.0/16{\textunderscore}013/0001402,
CZ.02.1.01/0.0/0.0/18{\textunderscore}046/0016010 and
CZ.02.1.01/0.0/0.0/17{\textunderscore}049/0008422; France -- Centre de Calcul
IN2P3/CNRS; Centre National de la Recherche Scientifique (CNRS); Conseil
R\'egional Ile-de-France; D\'epartement Physique Nucl\'eaire et Corpusculaire
(PNC-IN2P3/CNRS); D\'epartement Sciences de l'Univers (SDU-INSU/CNRS);
Institut Lagrange de Paris (ILP) Grant No.~LABEX ANR-10-LABX-63 within
the Investissements d'Avenir Programme Grant No.~ANR-11-IDEX-0004-02;
Germany -- Bundesministerium f\"ur Bildung und Forschung (BMBF); Deutsche
Forschungsgemeinschaft (DFG); Finanzministerium Baden-W\"urttemberg;
Helmholtz Alliance for Astroparticle Physics (HAP);
Helmholtz-Gemeinschaft Deutscher Forschungszentren (HGF); Ministerium
f\"ur Innovation, Wissenschaft und Forschung des Landes
Nordrhein-Westfalen; Ministerium f\"ur Wissenschaft, Forschung und Kunst
des Landes Baden-W\"urttemberg; Italy -- Istituto Nazionale di Fisica
Nucleare (INFN); Istituto Nazionale di Astrofisica (INAF); Ministero
dell'Istruzione, dell'Universit\'a e della Ricerca (MIUR); CETEMPS Center
of Excellence; Ministero degli Affari Esteri (MAE); M\'exico -- Consejo
Nacional de Ciencia y Tecnolog\'\i{}a (CONACYT) No.~167733; Universidad
Nacional Aut\'onoma de M\'exico (UNAM); PAPIIT DGAPA-UNAM; The Netherlands
-- Ministry of Education, Culture and Science; Netherlands Organisation
for Scientific Research (NWO); Dutch national e-infrastructure with the
support of SURF Cooperative; Poland -Ministry of Science and Higher
Education, grant No.~DIR/WK/2018/11; National Science Centre, Grants
No.~2013/08/M/ST9/00322, No.~2016/23/B/ST9/01635 and No.~HARMONIA
5--2013/10/M/ST9/00062, UMO-2016/22/M/ST9/00198; Portugal -- Portuguese
national funds and FEDER funds within Programa Operacional Factores de
Competitividade through Funda\c{c}\~ao para a Ci\^encia e a Tecnologia
(COMPETE); Romania -- Romanian Ministry of Education and Research, the
Program Nucleu within MCI (PN19150201/16N/2019 and PN19060102) and
project PN-III-P1-1.2-PCCDI-2017-0839/19PCCDI/2018 within PNCDI III;
Slovenia -- Slovenian Research Agency, grants P1-0031, P1-0385, I0-0033,
N1-0111; Spain -- Ministerio de Econom\'\i{}a, Industria y Competitividad
(FPA2017-85114-P and FPA2017-85197-P), Xunta de Galicia (ED431C
2017/07), Junta de Andaluc\'\i{}a (SOMM17/6104/UGR), Feder Funds, RENATA Red
Nacional Tem\'atica de Astropart\'\i{}culas (FPA2015-68783-REDT) and Mar\'\i{}a de
Maeztu Unit of Excellence (MDM-2016-0692); USA -- Department of Energy,
Contracts No.~DE-AC02-07CH11359, No.~DE-FR02-04ER41300,
No.~DE-FG02-99ER41107 and No.~DE-SC0011689; National Science Foundation,
Grant No.~0450696; The Grainger Foundation; Marie Curie-IRSES/EPLANET;
European Particle Physics Latin American Network; and UNESCO.
\end{sloppypar}


\bibliographystyle{aasjournal}

\newpage

\AuthorCollaborationLimit=3000
{\bf\Large{The Pierre Auger Collaboration}}\\


A.~Aab$^{75}$,
P.~Abreu$^{67}$,
M.~Aglietta$^{50,49}$,
J.M.~Albury$^{12}$,
I.~Allekotte$^{1}$,
A.~Almela$^{8,11}$,
J.~Alvarez-Mu\~niz$^{74}$,
R.~Alves Batista$^{75}$,
G.A.~Anastasi$^{58,49}$,
L.~Anchordoqui$^{82}$
% ...


\begin{description}[labelsep=0.2em,align=right,labelwidth=0.7em,labelindent=0em,leftmargin=2em,noitemsep]
\item[$^{1}$] Centro At\'omico Bariloche and Instituto Balseiro (CNEA-UNCuyo-CONICET), San Carlos de Bariloche, Argentina
\item[$^{2}$] Centro de Investigaciones en L\'aseres y Aplicaciones, CITEDEF and CONICET, Villa Martelli, Argentina
\item[$^{3}$] Departamento de F\'\i{}sica and Departamento de Ciencias de la Atm\'osfera y los Oc\'eanos, FCEyN, Universidad de Buenos Aires and CONICET, Buenos Aires, Argentina
\item[$^{4}$] IFLP, Universidad Nacional de La Plata and CONICET, La Plata, Argentina
\item[$^{5}$] Instituto de Astronom\'\i{}a y F\'\i{}sica del Espacio (IAFE, CONICET-UBA), Buenos Aires, Argentina
\item[$^{6}$] Instituto de F\'\i{}sica de Rosario (IFIR) -- CONICET/U.N.R.\ and Facultad de Ciencias Bioqu\'\i{}micas y Farmac\'euticas U.N.R., Rosario, Argentina
\item[$^{7}$] Instituto de Tecnolog\'\i{}as en Detecci\'on y Astropart\'\i{}culas (CNEA, CONICET, UNSAM), and Universidad Tecnol\'ogica Nacional -- Facultad Regional Mendoza (CONICET/CNEA), Mendoza, Argentina
\item[$^{8}$] Instituto de Tecnolog\'\i{}as en Detecci\'on y Astropart\'\i{}culas (CNEA, CONICET, UNSAM), Buenos Aires, Argentina
\item[$^{9}$] Observatorio Pierre Auger, Malarg\"ue, Argentina
\item[$^{10}$] Observatorio Pierre Auger and Comisi\'on Nacional de Energ\'\i{}a At\'omica, Malarg\"ue, Argentina
\item[$^{11}$] Universidad Tecnol\'ogica Nacional -- Facultad Regional Buenos Aires, Buenos Aires, Argentina
\item[$^{12}$] University of Adelaide, Adelaide, S.A., Australia
\item[$^{13}$] Universit\'e Libre de Bruxelles (ULB), Brussels, Belgium
\item[$^{14}$] Vrije Universiteit Brussels, Brussels, Belgium
\item[$^{15}$] Centro Brasileiro de Pesquisas Fisicas, Rio de Janeiro, RJ, Brazil
\item[$^{16}$] Centro Federal de Educa\c{c}\~ao Tecnol\'ogica Celso Suckow da Fonseca, Nova Friburgo, Brazil
\item[$^{17}$] Universidade de S\~ao Paulo, Escola de Engenharia de Lorena, Lorena, SP, Brazil
\item[$^{18}$] Universidade de S\~ao Paulo, Instituto de F\'\i{}sica de S\~ao Carlos, S\~ao Carlos, SP, Brazil
\item[$^{19}$] Universidade de S\~ao Paulo, Instituto de F\'\i{}sica, S\~ao Paulo, SP, Brazil
\item[$^{20}$] Universidade Estadual de Campinas, IFGW, Campinas, SP, Brazil
\item[$^{21}$] Universidade Estadual de Feira de Santana, Feira de Santana, Brazil
\item[$^{22}$] Universidade Federal do ABC, Santo Andr\'e, SP, Brazil
\item[$^{23}$] Universidade Federal do Paran\'a, Setor Palotina, Palotina, Brazil
\item[$^{24}$] Universidade Federal do Rio de Janeiro, Instituto de F\'\i{}sica, Rio de Janeiro, RJ, Brazil
\item[$^{25}$] Universidade Federal do Rio de Janeiro (UFRJ), Observat\'orio do Valongo, Rio de Janeiro, RJ, Brazil
\item[$^{26}$] Universidade Federal Fluminense, EEIMVR, Volta Redonda, RJ, Brazil
\item[$^{27}$] Universidad de Medell\'\i{}n, Medell\'\i{}n, Colombia
\item[$^{28}$] Universidad Industrial de Santander, Bucaramanga, Colombia
\item[$^{29}$] Charles University, Faculty of Mathematics and Physics, Institute of Particle and Nuclear Physics, Prague, Czech Republic
\item[$^{30}$] Institute of Physics of the Czech Academy of Sciences, Prague, Czech Republic
\item[$^{31}$] Palacky University, RCPTM, Olomouc, Czech Republic
\item[$^{32}$] Universit\'e Paris-Saclay, CNRS/IN2P3, IJCLab, Orsay, France, France
\item[$^{33}$] Laboratoire de Physique Nucl\'eaire et de Hautes Energies (LPNHE), Universit\'es Paris 6 et Paris 7, CNRS-IN2P3, Paris, France
\item[$^{34}$] Univ.\ Grenoble Alpes, CNRS, Grenoble Institute of Engineering Univ.\ Grenoble Alpes, LPSC-IN2P3, 38000 Grenoble, France, France
\item[$^{35}$] Bergische Universit\"at Wuppertal, Department of Physics, Wuppertal, Germany
\item[$^{36}$] Karlsruhe Institute of Technology, Institute for Experimental Particle Physics (ETP), Karlsruhe, Germany
\item[$^{37}$] Karlsruhe Institute of Technology, Institut f\"ur Kernphysik, Karlsruhe, Germany
\item[$^{38}$] Karlsruhe Institute of Technology, Institut f\"ur Prozessdatenverarbeitung und Elektronik, Karlsruhe, Germany
\item[$^{39}$] RWTH Aachen University, III.\ Physikalisches Institut A, Aachen, Germany
\item[$^{40}$] Universit\"at Hamburg, II.\ Institut f\"ur Theoretische Physik, Hamburg, Germany
\item[$^{41}$] Universit\"at Siegen, Fachbereich 7 Physik -- Experimentelle Teilchenphysik, Siegen, Germany
\item[$^{42}$] Gran Sasso Science Institute, L'Aquila, Italy
\item[$^{43}$] INFN Laboratori Nazionali del Gran Sasso, Assergi (L'Aquila), Italy
\item[$^{44}$] INFN, Sezione di Catania, Catania, Italy
\item[$^{45}$] INFN, Sezione di Lecce, Lecce, Italy
\item[$^{46}$] INFN, Sezione di Milano, Milano, Italy
\item[$^{47}$] INFN, Sezione di Napoli, Napoli, Italy
\item[$^{48}$] INFN, Sezione di Roma ``Tor Vergata'', Roma, Italy
\item[$^{49}$] INFN, Sezione di Torino, Torino, Italy
\item[$^{50}$] Osservatorio Astrofisico di Torino (INAF), Torino, Italy
\item[$^{51}$] Politecnico di Milano, Dipartimento di Scienze e Tecnologie Aerospaziali , Milano, Italy
\item[$^{52}$] Universit\`a del Salento, Dipartimento di Matematica e Fisica ``E.\ De Giorgi'', Lecce, Italy
\item[$^{53}$] Universit\`a dell'Aquila, Dipartimento di Scienze Fisiche e Chimiche, L'Aquila, Italy
\item[$^{54}$] Universit\`a di Catania, Dipartimento di Fisica e Astronomia, Catania, Italy
\item[$^{55}$] Universit\`a di Milano, Dipartimento di Fisica, Milano, Italy
\item[$^{56}$] Universit\`a di Napoli ``Federico II'', Dipartimento di Fisica ``Ettore Pancini'', Napoli, Italy
\item[$^{57}$] Universit\`a di Roma ``Tor Vergata'', Dipartimento di Fisica, Roma, Italy
\item[$^{58}$] Universit\`a Torino, Dipartimento di Fisica, Torino, Italy
\item[$^{59}$] Benem\'erita Universidad Aut\'onoma de Puebla, Puebla, M\'exico
\item[$^{60}$] Centro de Investigaci\'on y de Estudios Avanzados del IPN (CINVESTAV), M\'exico, D.F., M\'exico
\item[$^{61}$] Unidad Profesional Interdisciplinaria en Ingenier\'\i{}a y Tecnolog\'\i{}as Avanzadas del Instituto Polit\'ecnico Nacional (UPIITA-IPN), M\'exico, D.F., M\'exico
\item[$^{62}$] Universidad Aut\'onoma de Chiapas, Tuxtla Guti\'errez, Chiapas, M\'exico
\item[$^{63}$] Universidad Nacional Aut\'onoma de M\'exico, M\'exico, D.F., M\'exico
\item[$^{64}$] Institute of Nuclear Physics PAN, Krakow, Poland
\item[$^{65}$] University of \L{}\'od\'z, Faculty of Astrophysics, \L{}\'od\'z, Poland
\item[$^{66}$] University of \L{}\'od\'z, Faculty of High-Energy Astrophysics,\L{}\'od\'z, Poland
\item[$^{67}$] Laborat\'orio de Instrumenta\c{c}\~ao e F\'\i{}sica Experimental de Part\'\i{}culas -- LIP and Instituto Superior T\'ecnico -- IST, Universidade de Lisboa -- UL, Lisboa, Portugal
\item[$^{68}$] ``Horia Hulubei'' National Institute for Physics and Nuclear Engineering, Bucharest-Magurele, Romania
\item[$^{69}$] Institute of Space Science, Bucharest-Magurele, Romania
\item[$^{70}$] University Politehnica of Bucharest, Bucharest, Romania
\item[$^{71}$] Center for Astrophysics and Cosmology (CAC), University of Nova Gorica, Nova Gorica, Slovenia
\item[$^{72}$] Experimental Particle Physics Department, J.\ Stefan Institute, Ljubljana, Slovenia
\item[$^{73}$] Universidad de Granada and C.A.F.P.E., Granada, Spain
\item[$^{74}$] Instituto Galego de F\'\i{}sica de Altas Enerx\'\i{}as (IGFAE), Universidade de Santiago de Compostela, Santiago de Compostela, Spain
\item[$^{75}$] IMAPP, Radboud University Nijmegen, Nijmegen, The Netherlands
\item[$^{76}$] KVI -- Center for Advanced Radiation Technology, University of Groningen, Groningen, The Netherlands
\item[$^{77}$] Nationaal Instituut voor Kernfysica en Hoge Energie Fysica (NIKHEF), Science Park, Amsterdam, The Netherlands
\item[$^{78}$] Stichting Astronomisch Onderzoek in Nederland (ASTRON), Dwingeloo, The Netherlands
\item[$^{79}$] Universiteit van Amsterdam, Faculty of Science, Amsterdam, The Netherlands
\item[$^{80}$] Case Western Reserve University, Cleveland, OH, USA
\item[$^{81}$] Colorado School of Mines, Golden, CO, USA
\item[$^{82}$] Department of Physics and Astronomy, Lehman College, City University of New York, Bronx, NY, USA
\item[$^{83}$] Louisiana State University, Baton Rouge, LA, USA
\item[$^{84}$] Michigan Technological University, Houghton, MI, USA
\item[$^{85}$] New York University, New York, NY, USA
\item[$^{86}$] Pennsylvania State University, University Park, PA, USA
\item[$^{87}$] University of Chicago, Enrico Fermi Institute, Chicago, IL, USA
\item[$^{88}$] University of Delaware, Department of Physics and Astronomy, Bartol Research Institute, Newark, DE, USA
\item[] -----
\item[$^{a}$] School of Physics and Astronomy, University of Leeds, Leeds, United Kingdom
\item[$^{b}$] Max-Planck-Institut f\"ur Radioastronomie, Bonn, Germany
\item[$^{c}$] Fermi National Accelerator Laboratory, USA
\item[$^{d}$] also at Universidade Federal de Alfenas, Po\c{c}os de Caldas, Brazil
\item[$^{e}$] Colorado State University, Fort Collins, CO, USA
\item[$^{f}$] now at Hakubi Center for Advanced Research and Graduate School of Science, Kyoto University, Kyoto, Japan
\item[$^{g}$] also at Karlsruhe Institute of Technology, Karlsruhe, Germany
\item[$^{h}$] also at University of Bucharest, Bucharest, Romania
\end{description}

\end{document}